# Generic role of the anisotropic surface free energy on the morphological evolution in a strained-heteroepitaxial solid droplet on a rigid substrate

Tarik Omer Ogurtani,[1] Aytac Celik[2] and Ersin Emre Oren[3]

*Department of Metallurgical and Materials Engineering, Middle East Technical University, 06531, Ankara, Turkey*

## ABSTRACT

A systematic study based on the self-consistent dynamical simulations is presented for the spontaneous evolution of an isolated thin solid droplet on a rigid substrate, which is driven by the surface drift diffusion induced by the anisotropic capillary forces (surface stiffness) and mismatch stresses. In this work, we studied the affect of surface free energy anisotropies (weak and strong (anomalous)) on the development kinetics of the '*Stranski-Krastanow*' island type morphology. The anisotropic surface free energy and the surface stiffness were treated with well accepted trigonometric functions. Although, various tilt angles and anisotropy constants were considered during simulations, the main emphasis was given on the effect of rotational symmetries associated with the surface Helmholtz free energy topography in 2D space. Our computer simulations revealed the formation of an extremely thin wetting layer during the development of the bell-shaped Stranski-

---

[1] Corresponding author, Tel.: +90-312-210-2512; fax: +90-312-210-1267; e-mail: ogurtani@metu.edu.tr
 Url: http://www.csl.mete.metu.edu.tr

[2] Electronic mail: e104548@metu.edu.tr

[3] Electronic mail: eeoren@uw.edu

Krastanow island through the mass accumulation at the central region of the droplet via surface drift-diffusion. For weak anisotropy constant levels, instead of singlet islanding, we also observed formation of doublet islanding, separated by shallow wetting layer, for a set of specific tilt angles, $\phi = 90^o$ and $\phi = 45^o$, respectively, for the two-fold and four-fold rotational symmetry axis. No such formation has been detected for the six-fold symmetry. In the strong (anomalous) anisotropy constant domain, we demonstrated the existence of two distinct morphological modes: i) the *complete stability* of the initial Cosine-shaped droplet just above a certain anisotropy constant threshold level by spontaneous slight readjustments of the base and the height of the cluster; ii) the *Frank-van der Merwe* mode of thin film formation for very large values of the anisotropy constant by the *spreading* and *coalescence* of the droplets over the substrate surface. During the course of the simulations, we have continuously tracked both the morphology (*i.e.*, the peak height, the extension of the wetting layer beyond the domain boundaries, and the triple junction contact angle) and energetic (the global Helmholtz free energy changes associated with the total strain and surface energy variations) of the system.

## I. INTRODUCTION

The formation of 'nanoscale islands' or 'quantum dots (QDs)' separated by a thin flat wetting layer, known as Stranski-Krastanow (SK) morphology, is a general growth mode observed in many epitaxially-strained thin solid film systems such as in [$In_xGa_{1-x}$ As/GaAs][1] and [Ge/Si][2]. The formation of quantum dots through the SK growth mode on epitaxially strained thin film surfaces has attracted great attention over the last two decades due to their unique electronic and optical properties.[3,4] The fundamental understanding of the SK growth mode will provide insights necessary to control precise positioning, density and size of QDs and may open new avenues in QDs fabrication techniques. During the SK growth, the formation of QDs can be initiated via two diverse mechanisms: nucleation[5] or surface roughening (nucleationless),[6] followed by island growth and coarsening. The former usually gives rise to a wide range of island distribution since nucleation is a stochastic process, and thus it is imprecise. Whereas the latter process usually generates a relatively more uniform and regular island arrays via the cooperative formation of the QDs.

It was shown so far that there are various parameters including surface energy anisotropy, strain level, wetting conditions and growth kinetics, affecting how the surface evolution would reach a prescribed stationary state *i.e.* SK morphology. In general, nonlinear analyses in two dimensional configurations revealed that the stress-driven surface instabilities evolve into deep, crack like groove or cusp morphologies.[7] However, unlike the semi-infinite homogenous solids, the presence of a substrate affects the instabilities in several ways: First, a stiffer substrate tends to stabilize the film and increases the critical wave length, while the opposite is true for softer substrates. At the limit of a rigid substrate, a critical film thickness exists as shown by Spencer *et al.*,[8,9] below which the film is stable against perturbations of any wave lengths. Furthermore, the existence of an interface between the film and the substrate brings more complexity to the problem. At the close proximity of the film surface to substrate, short range wetting interactions dominate and cause an increase in the local surface free energy of the film. This increase hinders the penetration of islands through the boundary layer, and thereby avoids the formation of the *Volmer-Weber* (VW) type of growth mode (*i.e.*, island formation) and promote SK growth mode by preventing the surface of the substrate between islands from exposure to the immediate environment.[10,11,12]

Spencer[10] and Tekalign and Spencer[13,14] made extensive and very successful 2D and 3D analyses on the morphological instability of growing epitaxially strained dislocation-free solid films. Their analyses were based on the surface diffusion driven by the capillary forces and misfit strains by elaborating various type of wetting potentials associated with the thickness dependent surface specific free energy. In their work, similar to the simulation studies of the stability of epitaxially strained islands by Chiu and Gao,[11] Zhang and Bower,[15] Zhang,[16] Krishnamurthy and Srolovitz,[17] Medhekar and Shenoy,[18] and Levine *et al.*,[19] elastic strain energy density (ESED) appears to be additive to the generalized chemical potential. Zhang and Bower,[20] and Liu *et al.*[21] made a highly regarded 3D simulation studies on the shape transitions and coarsening in strained-heteroepitaxial islands. Including these two, the majority of the numerical and analytical studies, reported in the literature for the so-called steady state solutions of the nonlinear immobile-boundary value problem, utilized the periodic boundary conditions and relied mostly on the instabilities initiated by the white noise or the small amplitude initial perturbations, where the film thickness is smaller than the wavelength of surface variations.

The relationship between the film-substrate wetting interactions and faceting instability, caused by anisotropic surface free energy has been studied thoroughly by Golovin *et al*.[22] They showed that, even in the absence of epitaxial stresses, the wetting interactions can terminate coarsening and lead to the formation of permanent arrays of quantum dots, as well as spatially localized dots. In later studies, performed by Eisenberg and Kandel,[23] faceted island formation in films with anisotropic surface tension has been shown by using two-dimensional (2D) simulations. It has also been demonstrated that a single cusp in the surface free energy is sufficient to explain the observed island-shape transition and associated bimodal island size-distribution.

In the present work, we took into account the effects of not only the epitaxial stresses but also the surface free energy anisotropies fully via considering the tilt angles as well as the various *rotational fold-numbers* (*i.e.* n=2, 4,and 6) in cubic structures associated with *the stiffness map* in 2D space on the final morphologies. Our main study still focuses on the first route in which the islands forms through the nucleation by producing shallow droplets rather than surface roughening. The present simulations revealed many detailed picture of the shape transition and kinetics of the isolated-droplet towards the SK island morphology without the growth process in the absence of coarsening. Here, we also lifted the restriction of immobile-boundary conditions on the triple junction (TJ) motion by employing an irreversible thermodynamic connection obtained by using the internal entropy production (IEP) hypothesis.[24] IEP hypothesis furnishes the temporal velocity of the TJ singularity with the instantaneous values of the contact angle (*i.e.*, one sided dihedral angle) and the wetting parameter, which depends only on the specific surface Helmholtz free energies of the thin film, substrate and the interface between them. This free-moving boundary condition promotes *the static coarsening*, where islands undergo coalescence only when the neighboring islands touch to each other, in addition to the Ostwald ripening. We also demonstrated that this free moving boundary condition at TJ becomes very important, especially for the anomalous stiffness regime. In the anomalous stiffness regime, the complete stabilization of droplets by slight base length readjustment or the smooth layer structure formation by spreading (*i.e.*, the *Frank-van der Merwe* growth mode) have been observed as dominant scenarios, respectively, for the moderate or very high values of the anisotropy constants above the threshold level.

## II. PHYSICAL AND MATHEMATICAL MODELING

A continuum theory, based on the microdiscrete formulation of the irreversible thermodynamics of surfaces and interfaces, was extensively elaborated and applied by Ogurtani[24,23] and Ogurtani and Oren[25] for multi-component systems. This approach has been enlarged here by taking into account the film thickness dependent surface Helmholtz free energy (*i.e.*, the wetting effect) to study the evolution behavior of epitaxial thin-films. The explicit inclusion of the wetting effect/potential associated with the film/substrate interface becomes especially important to simulate the shape transition towards the *Stranski-Krastanow* type islands, which are embedded into the ultra-thin wetting layer background, by taking the initial nucleation route (*i.e.*, thin solid droplet formation) as a starting point rather than the usual surface roughening scheme.

### a. The governing equation for the surface drift-diffusion without growth term

The evolution kinetics of surfaces or interfacial layers of an *isochoric multi-phase system* may be described in terms of surface normal displacement velocities, $\bar{V}_{ord}$, by the following well-posed free moving boundary value problem in 2D space for ordinary points (*i.e.*, the generalized cylindrical surfaces in 3D space) using normalized and scaled parameters and variables, which are indicated with the bar signs. Similarly, the TJ longitudinal velocity $\bar{V}_{Long}$ associated with the natural motion of the droplet-substrate contour line may be given in terms of the wetting parameter $\lambda = [(f_s - f_{ds})/f_d]$, and the temporal one-sided dihedral or wetting contact angle $\theta_W$ as a dynamical variable. Here $f_s$ is the Helmholtz surface free energy of the substrate, and $f_{ds}$ is the interfacial free energy between the droplet and the substrate, and $f_d$ is the surface free energy of the thick solid film; $f_d \to f_{d/s}(\infty)$, where $f_{d/s}(y)$ is the height dependent surface Helmholtz free energy of the droplet. According to our adopted sign convention, the negative values of $\bar{V}_{ord}$ and $\bar{V}_{Long}$ correspond to the local expansion and/or growth of a droplet:

$$\bar{V}_{ord} = \frac{\partial}{\partial \bar{\ell}} \left[ \frac{\partial}{\partial \bar{\ell}} \left( -\Sigma (\bar{\sigma}_h)^2 + \bar{f}_{d/s}(\bar{y}) \hat{f}(\theta, \phi, m) \bar{\kappa} + \bar{\omega}(\bar{y}) \right) \right] \text{ (Ordinary points)} \quad (1)$$

and

$$\bar{V}_{Long} = -\bar{M}_{Long} \bar{\Omega}^{-1} \{\lambda - \cos(\theta_W)\} \quad \forall \ \lambda \geq 1 \text{ (Triple junction contour line)} \quad (2)$$

A careful examination of Eq. (1) shows that: it is almost identical *in physico-mathematical context* to the 2D version of the governing equation employed successfully by Zhang and Bower[20,21], and Liu *et al*.,[21,22] during the 3D simulations of the scenarios associated with the SK island growth modes. In the present study, the growth term, originated by the condensation and/or evaporation processes taking place at the interface between the droplet and the vapor phase, is completely omitted in order to simplify the problem.

Eq. (2) defines the in-plane displacement velocity of the contour line (*i.e.*, the TJ line shared by the droplet, substrate and vapor phases). The various terms appearing in Eq. (1) and (2) are described as follows: $\bar{\Omega}$ is the normalized atomic volume in the particle representation by assuming tentatively that the scaling length is in the range of 10 atomic spacing.[26] $\bar{M}_{Long}$ is the ratio of the mobility of the TJ, $\hat{M}_{Long}$, to the surface mobility, $\hat{M}_d$. $\bar{\kappa}$ is the normalized local curvature and is taken to be positive for a concave solid surface (troughs), and the positive direction of the surface displacement and the surface normal vector $\hat{n}$ are assumed to be towards the bulk (*i.e.*, droplet) phase, and implies the local shrinkage or evaporation processes.

In the governing equation, Eq. (1), the normalized hoop stress is denoted by $\bar{\sigma}_h \equiv Tr\bar{\sigma}$, where the dimensionless stress intensity parameter $\Sigma$ corresponds to the intensity of the ESED contribution on the stress-driven surface drift diffusion. The misfit strain $\varepsilon_o$ at the film/substrate interface is introduced as a *Dirichlet boundary condition* by specifying the displacement vector in 2D space as $\tilde{u} \rightarrow \hat{i}\varepsilon_o x$ (*i.e.*, in 3D pseudo-space $\tilde{u} \rightarrow \left[\hat{i}\varepsilon_o x, \hat{k}\varepsilon_o z\right]$), and taking the droplet center at the film/substrate interface as the origin of the coordinate system to avoid shifting. Similarly, the stress used for the normalization procedure is chosen as the biaxial stress $\sigma_o = E_d \varepsilon_o / (1 - v_d)$, where, $E_d$ and $v_d$ are, respectively, Young modulus and Poisson's ratio of the droplet shape film, and $\varepsilon_o$ is misfit strain at the film/substrate interface. Here, we assumed that the surface Helmholtz free energy density $f_{d/s}(y)$ for an isochoric system depends on the local distance $y$ between the surface layer and the substrate. In Eq. (1), $\bar{\omega}(y)$ is the normalized wetting potential, which is given by $\omega(y) = \Omega_d n_y df_{d/s}/dy$ in particle representation, where $n_y = -\hat{n}.\hat{j}$ is the projection of the surface normal along the $y$-axis, which is taken as perpendicular to the substrate. In the above expression, $\bar{\ell}$ is the curvilinear coordinate along the surface (arc length) in 2D space scaled with respect to $\ell_o$. Here, $\ell_o$ is the arbitrary length scale and may be selected as the peak height $h_p$ of the droplet or the ratio of the surface Helmholtz free energy of the film in the bulk to the ESED[11,14] such as $\ell^* = f_d / w_o$. Here, $w_o = (1 - v_d^2)\sigma_o^2 / 2E_d$ denotes ESED, which is associated with the nominal biaxial misfit stress taking the third dimension into account. If one takes the *characteristic length* denoted by $\ell^*$ as a scaling parameter then one should have the following replacement: $\Sigma \leftarrow 1$, because according to our definition $\Sigma \equiv \ell_o / \ell^*$. In the present study, otherwise it is stated, the initial peak height of the droplet $h_p$ is chosen as the natural scaling length, namely; $\ell_o = h_p$. The film thickness $h_o$ is defined as the *integrated film thickness,* and it may be given by $\bar{h}_o = 2\bar{h}_p / \pi \to 0.637$ for the scaled halve wave length Cosine-shape flat droplets, where $\bar{h}_p \to 1$.

In Eq. (1), $\hat{f}(\theta,\phi,m)$ corresponds to the angular part of the *surface stiffness*, given by $\hat{f}(\theta,\phi,m) = \bar{f}(\theta,\phi,m) + \bar{f}_{\theta\theta}(\theta,\phi,m)$, where $\bar{f}(\theta,\phi,m)$ is the normalized angular part of the surface Helmholtz free energy. By following the general trend, one may introduce the trigonometric representation by defining the tilt angle $\phi$ such as that the *surface tangent* of a selected vicinal plane coincides with the *x*-axis when $\phi$ becomes equal to zero, $\phi$=0. Hence, the initial flat surface represented by $\theta = 0$ corresponds to one of those minima or cusp in the surface Helmholtz free energy described by the Wulff construction in 2D space:

$$\bar{f}(\theta,\phi,m) = f(\theta,\phi,m) / f_o = \{1 + B\sin^2[m(\theta - \phi)]\} \quad (3)$$

Where, $f_o$ is the minimum value of the surface free energy density and $B \geq 0$ is the *anisotropy constant* which is a positive quantity in the above *ad hoc* representation and measures the fractional roughness on the Wulff construction of the surface free energy. Using above expression, one may deduce the surface stiffness as defined previously as:

$$\hat{f}(\theta,\phi,m) = f_o(1 + B/2)\left\{1 - \frac{B(1 - 4m^2)}{B + 2}\cos^2[2m(\theta - \phi)]\right\} \quad (4)$$

Here, $\theta$ is the angle between the *tangent line vector* of the diffusion plane of a generalized cylindrical surface projected onto 2D space and the *x*-axis of the global Cartesian reference system.

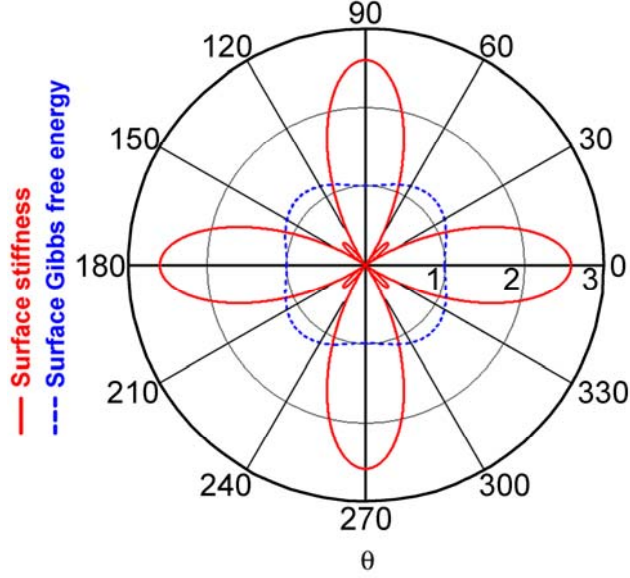

**FIG. 1. (Color online)** Typical behavior of surface specific free energy and surface stiffness for a set of four-fold symmetry planes $\{0\bar{1}0\}$ and $\{\pm 100\}$ in a fcc crystal having $[001]$ *zone axis* normal to the thin film surface. The anisotropy constant: $B = 0.2 > 1/7$. The negative spikes of the surface stiffness for the sidewalls drift-diffusion can clearly be seen along the $\langle 1\bar{1}0 \rangle$ directions, which indicates inherent anomalous instability.

In Fig. (1), the normalized specific surface Helmholtz free energy $\bar{f}$ and the angular part of the surface stiffness $\hat{f}$ are illustrated in the polar plot for *sidewall planes* of a thin film metallic single crystal, having a surface texture denoted by $(001)$ plane, for which $n \equiv 2m = 4$. According to Eq. (4), if the surface specific Helmholtz free energy anisotropy constant satisfies the following inequality: $B \leq 2/\left[\left|\left(1-4m^2\right)\right|-1\right]$ then the surface stiffness can be positive definite, regardless of the orientation of the surface with respect to the *x*-axis. That means one should have the following set of upper limits for the anisotropy constants $B \leq \{1; 1/7; 1/17\}$ in the case of two-, four- and six-fold symmetries, respectively. Otherwise, the system enters into the anomalous surface stiffness-induced instability regime. As can be seen from Fig. (1) for a given anisotropy constant $B = 0.2 > 1/7$ for the set of planes belonging to the $[001]$ *zone axis*, the imperfect faceting may occur at the cusp orientations (vicinal planes), because of the appreciable negative surface stiffness appears at the directions $\langle \overset{\pm}{110} \rangle$, where one has concave topography (maxima in free energy profile). These negative surface stiffness spikes may cause inherent anomalous instability along those directions as will be discussed later in this paper. In general, the factional variations in the anisotropy constant is in the range of $|\delta B/B| \leq 0.20$ for the series of vicinal planes $\{111\} < \{110\} < \{100\}$, for the most fcc and bcc metals and alloys.[27,28]

In the present enlarged formulation of the problem, we scaled the time and space variables $\{t, \ell\}$ in the following fashion: first of all, $\hat{M}_d$ an atomic mobility associated with the mass flow at the surface layer is defined, and then a normalized time scale is introduced by $\tau_o = \ell_o^4 / \left( \Omega^2 \hat{M}_d f_d \right)$. The bar signs over the letters indicate the following scaled and normalized quantities:

$$\bar{t}=t/\tau_o,\ \bar{\ell}=\ell/\ell_o,\ \bar{\kappa}=\kappa\,\ell_o,\ \bar{L}=L/\ell_o,\ \bar{\sigma}_h=\frac{\sigma_h}{\sigma_o} \tag{5}$$

$$w_o=\frac{(1-\nu_d^2)}{2E_d}\sigma_o^2,\ \sigma_o=\frac{E_d}{(1-\nu_d)}\varepsilon_o,\ \Sigma=\frac{(1-\nu_d^2)\ell_o}{2E_d f_d}(\sigma_o)^2\equiv\frac{\ell_o}{\ell^*} \tag{6}$$

$$\bar{\omega}(\bar{y})=\frac{1}{\sqrt{1+\bar{y}_x^2}}\frac{(f_s-f_d)}{\pi f_d}\frac{\bar{\delta}}{\bar{\delta}^2+\bar{y}^2} \tag{7}$$

and,

$$\bar{f}_{d/s}(y)=\frac{(f_d+f_s)}{2f_d}+\left(\frac{(f_d-f_s)}{f_d}\right)\frac{1}{\pi}\arctan(y/\delta) \tag{8}$$

Here, we adapted a transition-layer model as advocated by Spencer,[10] but reserved the case for the description of the wetting constant $\lambda$ since Spencer[10] and his coworkers[13] assumed that the interfacial free energy between the droplet and the substrate, $f_{ds}$, is negligible. According to the functional relationship given in Eq. (8) for the boundary layer model, the film specific Helmholtz surface free energy undergoes a rapid transition from film, $f_d$, to substrate, $f_s$, values over a length scale denoted as $\delta$. This roughly corresponds to the interfacial width and may be selected empirically in the range of $\delta\approx 0.10-0.01\ nm$.[13,29]

In the present study, the generalized mobility, $\bar{M}_{dv}$, associated with the interfacial displacement reaction (adsorption or desorption) is assumed to be independent of the orientation of the interfacial layer in crystalline solids. As we already mentioned, This generalized mobility is normalized with respect to the minimum value of the mobility of the surface drift-diffusion denoted by $\hat{M}_d$. They are given by: $\bar{M}_{dv} = (\hat{M}_{dv}\ell_o^2)/\hat{M}_d$ and, $\hat{M}_d = (\tilde{D}_d h_d / \bar{\Omega} kT)$ where, $\bar{\Omega}$ is the mean atomic volume of chemical species in the surface layer and $\tilde{D}_d$ is the isotropic part (*i.e.*, the minimum value in the case of anisotropy) of the surface diffusion coefficient.

**b. Implementation of the IBEM numerical method**

The detailed description of the indirect boundary elements method (IBEM), and its implementation[30] are presented very recently in two comprehensive papers by Ogurtani and Akyildiz[31,32] in connection with the void dynamics in metallic interconnects under the effect of electromigration forces. In this study, we utilized the simplest implementation of the IBEM that employs the straight constant line elements in the evaluation of the hoop stress at the free surface of the droplet, as well as along the interface between droplet and the substrate. In fact, it is also possible to generate the complete stress distribution field in the interior region of the sample as a byproduct. Here, Neumann (*i.e.*, traction free boundary condition) and Dirichlet boundary conditions (*i.e.*, prescribed displacements) are utilized, respectively, along the free surface of the droplet and at the interface between droplet and the substrate. Therefore, we implicitly assumed that the substrate is rigid, and the displacement is supplied as a Dirichlet boundary condition along with the interface, which is calculated from the misfit strain, $\varepsilon_o$, by $u_x(0) = \varepsilon_o x$. This implementation, adopted by the present author, guarantees the surface smoothness conditions for the validity of the governing *Fredholm integral equation* of the second kind at the corners and edges, which may be generated artificially during the numerical procedure. The explicit Euler's method combined with the adapted time step auto-control mechanism is employed in connection with *Gear's stiff stable* second-order time integration scheme[33] with the initial time step selected in the range of $\left(10^{-8}-10^{-9}\right)$ in the normalized time domain. This so-called adapted time step procedure combined with the self-recovery effect of the capillary terms guarantees the long-time numerical stability and accuracy of the explicit algorithm even after performing $2^{45}-2^{50}$ steps, which is clearly demonstrated in our recent work on the grain boundary grooving and cathode drifting in the presence of electromigration forces.[31,30] The network remeshing is continuously applied using the criteria advocated by Pan and Cocks,[34] and the curvature and normal line vector are evaluated at each node using a discrete geometric relationship in connection with the fundamental definitions of the radius of curvature.

## III. RESULTS AND DISCUSSION

In our simulations, the droplet is physically attached to the substrate with a coherent interface. The top surface is subjected to the surface drift diffusion, and also exposed to a vapor environment, whose pressure may be neglected. Since we are performing 2D simulations (equivalent to parallel ridges or quantum wires in three dimensions), no variation of the interface profile and the displacement fields in the droplet and substrate occurs in the direction (*i.e.*, $\hat{z}$ axis) perpendicular to the plane of the schematics in Fig. 2(a) (*i.e.*, plane strain condition). The symmetry axis, which is designated as the zone axis in this study is assumed to be laying down along the $\hat{z}$-axis of the substrate surface, which is span by the $\hat{x}$ and $\hat{z}$ sub-coordinate system.

In our computer simulation studies, it is further assumed that this thin crystalline droplet (*i.e.*, bump) initiated by the nucleation process may be described by a symmetrically disposed, halve-wave length *Cosine-function,* which is similar to the one employed by Kukta and Freund.[35] This initial configuration have a wave length and a height (*i.e.*, amplitude) denoted by 2*L* and $h_p$, respectively. The droplet *aspect ratio* may be defined by: $\beta = L/h_p$, which prescribes a finite contact angle $\theta = \arctan(\pi/\beta)$ between the film and the substrate at the onset of the simulation run. Therefore, in the normalized and scaled time-length space, the *initial shape* of a droplet is uniquely described by one single parameter, namely the aspect ratio $\beta$, since $\bar{h}_p = 1$ according the scheme we adopted in this study. Similarly, a close inspection of the normalized governing equation shows that there is only one more additional initial parameter left to complete the predetermination of the morphological evolution process, namely the elastic strain energy density, ESED parameter denoted by $\Sigma = \ell_o/\ell^* \to h_p/\ell^*$. The unitless $\Sigma$ parameter completely dictates the possible *size/area effects* of the droplet on the evolution process in *normalized space,* since *the characteristic length* $\ell^*$ is fixed by the physico-chemical properties of the system. On the other hand, the initial peak height $h_p = \ell_o = \Sigma\ell^*$ describes the size (*i.e.*, area) of the droplet in *real space* for a given aspect ratio $\beta$, namely: $A = 2\beta h_p^2/\pi$. Then, one may have the following expression for the normalized size for the droplet having the prescribed shape: $A_c/\ell^{*2} = 2\beta\,\Sigma^2/\pi$.

During the evolution process, the shape of the surface profile changes continuously. Hence, one has to utilize the power dissipation concept to calculate the global strain energy change of the droplet. This can be done by taking the time derivative of the bulk Helmholtz free energy (*i.e.*, the total strain energy) variation for an infinitesimal displacement of the surface layer along the surface normal designated as $\delta\eta$, which is given by the relationship $\delta W(\eta) = -\int_{Surf.} (w_d)\delta\eta d\ell$ for the traction free surfaces. According to our adopted definition of the surface normal, which is directed towards the solid phase, the shrinkage and the expansion of the solid phase is respectively corresponds to the inequalities $\delta\eta = -\delta h > 0$ and $\delta\eta = -\delta h < 0$. These results are in complete accord with findings by Rice and Drucker[36] and Gao[37] for the traction free surfaces. In a more general case, Eshelby[38] found a similar expression wich has an additional term related to *the energy-momentum tensor* for the bimaterial interfaces, which may carry non-vanishing tractions.

Here it has been presumed that the interface between film and substrate is immobile along the rigid substrate surface. Then one writes the following expression in the normalized space-time domain, where '$n$' designates the total number of nodes (collocation points) along the traction free surface, $\dot{\eta}$ indicates the total time differentiation, and $\bar{L}(\bar{t})$ is the instantaneous length of the free surface contour.

$$\bar{P}(\bar{t}) = -\int_{Surf.} (w^b)\dot{\eta}d\ell \Rightarrow -w_o \bar{L}(\bar{t}) \sum_{j=0}^{n-1} \frac{(\bar{\sigma}_j(\bar{t}))^2}{n} \dot{\eta}_j(\bar{t}). \qquad (9)$$

In Eq. (9), we excluded the contribution associated with the time variations in the strain energy density distribution evaluated at the free surface. Since, the numerical calculations showed that it is three orders of magnitude smaller than the one presented above with the same trend, *i.e.*, both having a negative sign. Subsequently, the cumulative change $\Delta \bar{F}_d$ in the *bulk Helmholtz* free energy, which is equal to the total elastic strain energy for the isothermal changes, during the evolution process may be calculated as a function of the discrete normalized time $\bar{t}_m$ by using a simple integration (*i.e.*, summation) procedure applied to above expression. This procedure yields:

$$\Delta \bar{F}_d \left( \bar{t}_m \right) \equiv \Delta W(\bar{t}_m) = \int_0^{\bar{t}_m} d\bar{t}\, \bar{P}(\bar{t}) \Rightarrow \bar{t}_m \sum_{k=0}^{k=m} \bar{P}(\bar{t}_k)/m. \tag{10}$$

Similarly, one may also write the Helmholtz surface free energy change $\Delta F_s$ associated with the free surface contour enlargement with respect to the initial configuration, which may be given for a prescribed time, $\bar{t}$, as;

$$\Delta \bar{F}_s \left( \bar{t} \right) = f_d \left[ \bar{L}(\bar{t}) - \bar{L}(0) \right] = w_o \ell^* \left[ \bar{L}(\bar{t}) - \bar{L}(0) \right] \tag{11}$$

In this paper, the bulk and surface Helmholtz free energy plots are normalized with respect to the nominal strain energy density $w_o = (1+\nu) E \varepsilon_o^2 /2$ to compare them properly even in the normalized space-time domain. For the future application in Ge/Si (100) system, the nominal elastic strain energy density may be given by $w_o \cong 1.58 \times 10^8\ N/M^2$. Our computer simulations show that one always observes the fulfillment of the following inequalities during the spontaneous evolution processes, $\Delta \bar{F}_d \left( \bar{t} \right) <0$ and $\Delta \bar{F}_s \left( \bar{t} \right) >0$. Even though their straight summation in normalized space may not be negative, one still expects for the natural isothermal processes occurring in the isochoric systems, the inequality $\Delta F_d(\bar{t}) + \Delta F_s(\bar{t}) < 0$ should be satisfied in the real time and length space. In fact, this requirement is also found to be satisfied in all the computer simulation experiments presented in this paper. Since the numerical calculations are carried out in normalized and scaled space, the following connections between the normalized and the real Helmholtz free energies associated with the elastic strain and the surface free energy contributions become very important. One may obtain these connections using the dimensional analysis as: $\Delta F_d = \ell_o^2 \Delta \bar{F}_d$ and $\Delta F_s = \ell_o \Delta \bar{F}_s$. Finally, these expression may be converted into following forms:

$$\Delta F_d(t) = \Sigma^2 \ell^{*2} \Delta \bar{F}_d(\bar{t}) \quad \text{and} \quad \Delta F_s(t) = \Sigma \ell^* \Delta \bar{F}_s(\bar{t}) \tag{12}$$

In the following sections, we will present the results obtained from a set of special computer experiments carried out on the specimens having large aspect ratios (*i.e.*, in the range of $\beta = (10-28)$), and subjected to the misfit strain at the interface between the thin film and the stiff substrate for various crystallographic surfaces in cubic structures. The surface of the droplet film initially is represented by a symmetrically disposed halve-wave length Cosine-curve as illustrated in Fig. 1(a), which has a normalized base length of $\beta = \bar{L} = 28$, and an amplitude of $\bar{h}_p = 1.0$ as compared to the *integrated film thickness*, which is given by $\bar{h}_o = 2\bar{h}_p / \pi \to 0.637$. Although, we employed a large number of different elastic strain energy density parameters (ESED) in our experiments, we will report only the one, which represents the formation of the Stranski-Krastanow singlet islanding in the cases of the isotropic and/or low anisotropic surface free energies, namely: $\Sigma = 0.40$. To illustrate the actual physical size of the islands, we consider the following parameters,[39] which are representative of Ge films grown epitaxially on a stiff silicon substrates.[35] Namely: $\varepsilon_o = -0.042$, $E_{Ge} = 103$ GPa, $\nu_{Ge} = 0.273$, $f_{Ge} = 1.927\ Jm^{-2}$, and $f_{Si} = 2.513\ Jm^{-2}$. These numbers entail a characteristic length of $\ell^* = 12.11\ nm$, which may be used to calculate the heights and the base lengths of the droplets that are corresponding to the range of the strain energy intensity parameters for a given aspect ratio (*i.e.*, $\beta = 28$), namely; for the singlet islands one has: $\{h_o = 3.08\ nm\} \cap \{L = 134\ nm\}$. Such sizes are typical of those observed experimentally by Krishnamurthy *et al.*[40]

### a. Two-fold rotational symmetry in cubic structures: Zone axis $\langle 1\bar{1}0 \rangle$

In this section, the results of the extensive computer simulation experiments performed on the single crystal thin film droplets attached to the substrate top surfaces described by two-fold rotational symmetry designated by the *zone-axis* $\langle 1\bar{1}0 \rangle$ are presented. This zone axis is associated with the set of vicinal (singular) planes such as $(111) \cup (110) \cup (001) \cup (11\bar{2})$ in cubic crystal structures. In the present study, we assume that $\{111\}$ *crystallographic planes* have the lowest surface free energy as observed experimentally[27,28] in fcc and bcc metals and alloys. Therefore zero tilt angle, which designates the orientation of the cusp in the surface free energy map in 2D space is associated with the (111) plane in the present simulation studies. The tilt angles, $\phi$, which are in the range of $(0^o - 180^o)$ are specially selected for demonstration purposes.

In Fig. 2, a typical morphological evolution behavior of the SK island deposited on the top of the Ge/Si$(111)$ vicinal plane is presented in terms of the final droplet profile. Here the peak height, the base extension, and the TJ contact angle are given with respect to the normalized logarithmic scale.

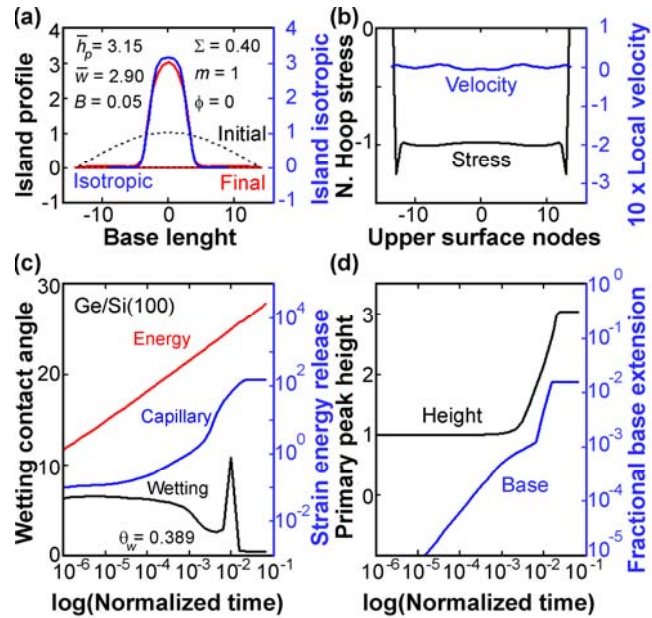

**FIG. 2. (Color online)** (a) Spontaneous formation of the SK island from a single crystal droplet on a stiff substrate via the surface drift diffusion driven by the combined actions of the misfit strain (isochoric) and the anisotropic capillary forces. (b) Instantaneous velocity and the hoop stress distributions along the final droplet profile. (c) Evolution of the contact angle is shown on the left *y*-axis. On the right *y*-axis, the strain energy and surface free energy changes are given for the Ge/Si(111) system. (d) Time evolution of peak height and TJ displacement. Data: $\Sigma = 0.40$, $B = 0.05$, $\phi = 0^o$ $\bar{h}_p = 1$, $\beta=28$, $\nu=0.273$, $\bar{M}_{TJ} = 2$, $\lambda=1$, $\bar{\delta} = 0.005$, $f_s = 1.2$ and $f_d = 1$.

In this experiment, we utilized a surface free energy anisotropy constant of $B = 0.05$ in connection with the elastic strain energy density parameter of $\Sigma = 0.40$, which was picked out from the upper bound of the stable singlet SK island formation range $\{\Sigma : 0.30, 0.35, 0.40\}$. This selected value of the anisotropy constant is well below the anomalous stiffness threshold level, which is previously given by $B_A = 1$. The SK profile reported in this figure shows a very thin simultaneously-formed wetting layer having a normalized thickness of $\Delta \bar{h} = 0.026$. This wetting layer thickness is about a factor of 5 greater than the adopted boundary layer thickness in our computer simulations, which enters as $\bar{\delta} = 0.005$ into the wetting potential presented in Eq. (1). In real space, the wetting layer thickness for the Ge/Si(111) system may be computed as follows: $\Delta h = 0.026 \ell_o \rightarrow 0.026 \Sigma \ell^* = 0.126 \ nm$ (*i.e.*, about one monolayer). The peak height of the SK island is found to be $\bar{h}_{SK} \cong 3.03 \rightarrow 14.53 \ nm$, which is slightly less than the height of isotropic SK island.

Fig. 2(c) demonstrates evolution behavior of the wetting layer at the TJ contour line, which has a temporal contact angle of $\theta = 0.74^o$ instead of zero degree, which indicates that the TJ is still active. In Fig. 2(c), the cumulative strain energy change, $-(\Delta F_d / w_o) nm^2$, as well as the increase in the surface free energy, $(\Delta F_s / w_o) nm^2$, of the droplet due to the island formation are also presented. Here the strain energy reduction shows a linear dependence on the normalized elapse time up to the onset of the stationary state regime. This fact may be easily anticipated by looking at the relevant plot, which with the exception of the initial transient stage has a slope of unity in the double logarithmic scale. A close inspection of Fig. 2(d) shows that the TJ displacement motion associated with the base extension has three different time exponent stages, $\bar{L}(\bar{t}) = A\bar{t}^n$ where $n = 1$; $\frac{1}{2}$; 6, before it enters to the plateau region. Similarly, the peak height shows a logarithmic time dependence during the intermediate regime before the onset of the plateau region, namely; $\bar{h}_p(\bar{t}) = 2\log(\bar{t}) + 5.6$.

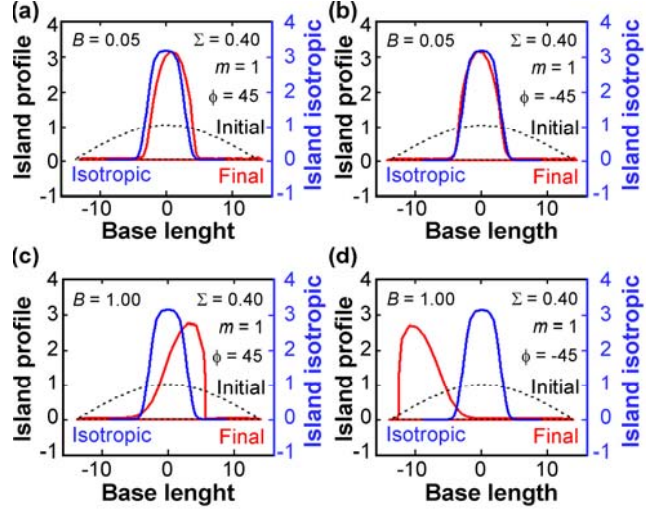

**FIG. 3. (Color online)** Spontaneous formations of the right- and left- shifted SK islands from single crystal droplets for (a and b) very low, $B = 0.05$, and (d and c) very high, $B = 1.0$, surface free energy anisotropy constants at $\pm 45^o$ tilt angles. At the threshold level of the anomalous instability regime, very sharp faceting at the right and left edges may be easily seen. Data: $\Sigma = 0.40$, $\bar{h}_p = 1$, $\beta = 28$, $\nu = 0.273$, $\bar{M}_{TJ} = 2$, $\lambda = 1$, $\bar{\delta} = 0.005$, $f_s = 1.2$ and $f_d = 1$.

In Fig. 3(a-d), we illustrate fully developed SK singlet islands, developed over those top substrate surfaces having $\phi = \pm 45^o$ tilt angles with respect to a member of the vicinal $\{111\}$ form of planes, which are assumed to be having smallest surface free energy.

The selected tilt angle $\phi = \pm 45^o$ is very close to the angle between (111) and (110) planes belonging to the $\langle 1\bar{1}0 \rangle$ zone axis, namely $\phi \cong 35.26^o$. Therefore, the results obtained here may easily be extrapolated to the SK islanding taking place in Ge/Si(110) system. Comparison between Fig. 3(a,b) and Fig. 3(c,d) shows that there is a substantial drop in the final SK peak height, and the large shift in the peak position if one goes from very low value of $B = 0.05$ to the high anisotropy constants of $B = 1.0$ even though the system still stays in the normal *Asaro-Tiller-Grinfeld* (ATG) instability regime. From the SK profiles reported in Fig. 3(c,d), one may easily see that the very sharp faceting with $90^o$ inclination to the platform is taking place on the right- and left-shoulders, respectively. The faceting planes may be identified as $(\bar{1}10)$ vicinal plane, which also belongs to the *zone-axis* $\langle 1\bar{1}0 \rangle$. At the stationary state regime, this faceted SK island is separated from the substrate by a thin wetting layer having a thickness of $\Delta \bar{h} = 0.030 \rightarrow 0.17\ nm$. In these last two cases, we utilized an ESED parameter of $\Sigma = 0.40$, and a surface free energy anisotropy constant of $B = 1.0$, which is just at the onset of the anomalous instability regime. The peak height and the halve width for the low anisotropy constant $B = 1.0$ found to be about $\bar{h}_{SK} = 2.74 \rightarrow 13.27\ nm$ and $\bar{w} = 2.06 \rightarrow 9.8\ nm$, respectively, compared to the one obtained for $B = 0.05$, namely: $\bar{h}_{SK} \cong 3.03 \rightarrow 14.53\ nm$.

In Fig. 4, we illustrate a fully developed SK *doublet* at the stationary state separated by a thin wetting layer having a thickness of $\Delta \bar{h} = 0.0314 \rightarrow 0.17\ nm$. The wetting layer thickness between the peaks, and the peak tails are found to be almost same in magnitude. In this case, we utilized an ESED parameter of $\Sigma = 0.40$, and a rather weak anisotropy constant of $B = 0.05$, with a tilt angle of $\phi = 90^o$. This tilt angle corresponds to $(11\bar{2})$ vicinal plane, which belongs to $\langle 1\bar{1}0 \rangle$ zone axis, and it is normal to the $(111)$ plane.

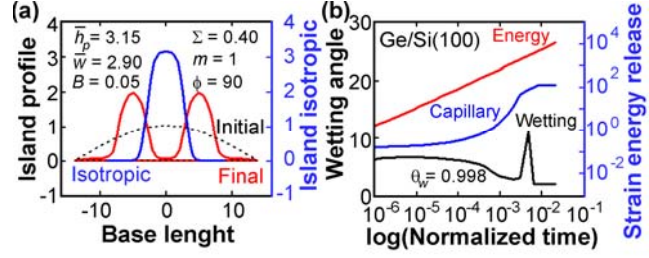

**FIG. 4. (Color online)** (a) Spontaneous formation of the SK doublets with an almost perfect flat wetting layer from a single crystal droplet having exposed to the anisotropic surface stiffness with a tilt angle of $\phi = 90^o$. (b) Evolution of the contact angle is shown on the left *y*-axis. On the right *y*-axis, the strain energy and surface free energy changes are given for Ge/Si$(11\bar{2})$ system. Data: $\Sigma = 0.40$, $B = 0.05$, $\phi = 90^o$ $\bar{h}_p = 1$, $\beta = 28$, $\nu = 0.273$, $\bar{M}_{TJ} = 2$, $\lambda = 1$, $\bar{\delta} = 0.005$, $f_s = 1.2$ and $f_d = 1$.

The extended plateau in the TJ wetting angle plot in Fig. 4(b) indicates that the stationary state equilibrium contact angle may not be necessarily realized, which should be otherwise equal to zero degree. These doublet peaks may be represented by the fourth degrees Gaussian type function $G(x; \bar{h}_p, \bar{w}) = \bar{h}_p \exp(-\ln(2) x^4 / \bar{w}^4)$, where the peak height and the halve width found to be $\bar{h}_{SK} = 1.98 \rightarrow 9.59\ nm$ and $\bar{w} = 2.06 \rightarrow 11.12\ nm$, respectively

**b. Four-fold rotational symmetry in cubic structures: Zone axis <001>**

In this section, we performed our computer simulations on single crystal thin film droplets having $(100)\cup(2\bar{1}0)\cup(1\bar{1}0)$ top substrate surfaces belonging to the $[001]$ zone axis, characterized by the four-fold rotational symmetry. In the first set of simulation experiments, a very low anisotropy constant such as $B = 0.05$ is selected in connection with the various tilt angles in the range of $\phi \supset (0^o - 90^o)$ to see the orientation effects on the morphological evolutions in the SK islanding. In the second set of experiments, the effects of the surface free energy anisotropy constant, which is chosen below and above the anomalous threshold level of $B_{th} = 1/7$ are examined in great details for the special tilt angle of $\phi = 0^o$. This orientation, which corresponds to the cusp in the Wulff free energy mapping is assumed to coincide with the top $(1\bar{1}0)$ surface of the substrate. This assumption as justified for fcc and bcc metals and alloys[27,28] implies that $(1\bar{1}0)$ plane has the lowest surface free energy among the all vicinal surfaces belonging to the $[001]$ zone axis. A typical top plane of a substrate that has a practical interest in Ge/Si(100) system may be described by any member of the form of planes $\{100\}$ as illustrated in Fig. 5.

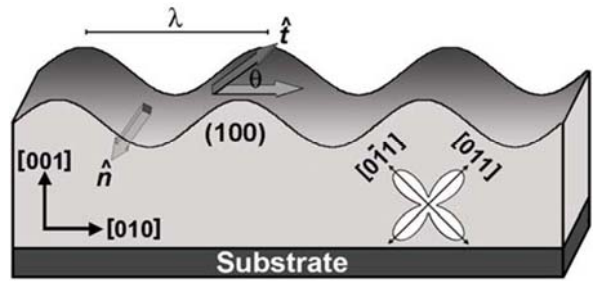

**FIG. 5.** Side view of a piece of metallic single crystal thin film, with a sinusoidal wave on the top surface of the wetting layer. This surface undulation is developed during evolution process in the *anomalous regime* because of the very high stiffness constant. This configuration corresponds to the four-fold rotation symmetry designated by $[100]$ zone axis. Where $n = 2m = 4$ and the tilt angle $\phi = 45^o$ if one takes the $\{1\bar{1}0\}$ form has lower free energy than the form of planes $\{100\}$, as far as the diffusion and the specific surface Helmholtz free energy dyadics are concerned.

To investigate the effects of the tilt angle on the morphological evaluation of a droplet we carried out experiments, where the test modulo is assumed to be exposed to an elastic strain energy density of $\Sigma = 0.4$, and the selected surface stiffness anisotropy constant $B = 0.05 \leq 1/7$ is below the threshold level of the anomalous regime. At the zero tilt angle case, these parameters result in a well developed SK singlet islanding having exactly the same shape parameters as reported in Fig. 2(a).

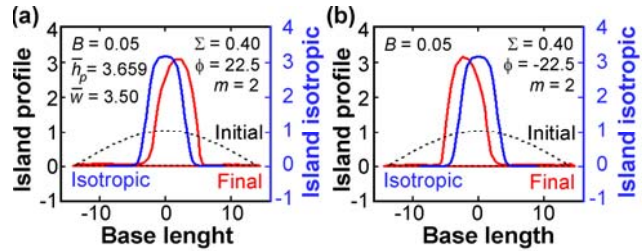

**FIG. 6. (Color online)** (a) Spontaneous formation of a tilted SK singlet towards the right edge with an almost perfect flat wetting layer from a single crystal droplet having exposed to the anisotropic surface stiffness with tilt angle of $\phi = 22.5^o$. (b) Where the same droplet is exposed to a tilt angle $\phi = -22.5^o$ by keeping all other system parameters same. Data: $\Sigma = 0.40$, $B = 0.05$, $\bar{h}_p = 1$, $\beta=28$, $\nu=0.273$, $\bar{M}_{TJ} = 2$, $\lambda=1$, $\bar{\delta} = 0.005$, $f_s = 1.2$ and $f_d = 1$.

In Fig. 6, the results obtained from a simulation experiments performed on a sample oriented with two different tilt angles, namely $\phi = 22.5^o$ and $\phi = -22.5^o$, are presented. These orientations almost correspond to the set of vicinal planes $(\bar{1}20) \cup (2\bar{1}0)$ symmetrically disposed with respect to the $(\bar{1}10)$ plane with the tilt angles of $\phi \cong \pm 18.4^o$. Fig. 6(a-b) clearly show the asymmetric morphologies associated with the SK singlet islands, having slightly tilted towards the right and the left edge, respectively for the tilt angles of $\phi = 22.5^o$ and $\phi = -22.5^o$. The kinetics output data, which are not reported here resembles those reported in Fig.3(c-d), and indicates that the surface free energy has reached to a stationary state, and TJ-contact angle is almost zero. On the contrary, the elastic strain energy release shows linear increase having a slope of unity on a double-logarithmic scale. That means there is no saturation in the energy release rate rather than it is a constant of time up to the onset of the stationary non-equilibrium SK-state.

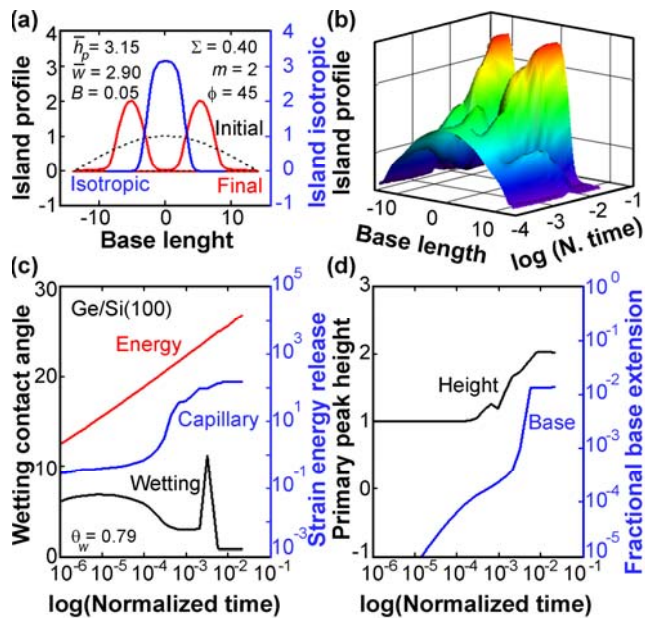

**FIG. 7. (Color online)** (a,b) Spontaneous formation of the SK doublets with an almost perfect flat wetting layer from a single crystal droplet having exposed to the anisotropic surface stiffness with tilt angle of $\phi = 45^o$. (c) Evolution of the contact angle is shown on the left *y*-axis. On the right *y*-axis, the strain energy and surface free energy changes are given for Ge/Si(100) system. (d) Time evolution of peak height and TJ displacement. Data: $\Sigma = 0.40$, $B = 0.05$, $\bar{h}_p = 1$, $\beta = 28$, $\nu = 0.273$, $\bar{M}_{TJ} = 2$, $\lambda = 1$, $\bar{\delta} = 0.005$, $f_s = 1.2$ and $f_d = 1$.

In Fig. 7(a,b) one can clearly see the morphology associated with the SK doublet islanding for the tilt angle of $\phi = 45^o$, which corresponds to the (100) top plane as illustrated in Fig. 5. The kinetics data, which is reported in Fig. 7(c) indicates that the surface free energy has reached to a stationary state, and TJ-contact angle is almost zero. Similarly, the elastic strain energy released shows linear increase with the normalized elapse time on a semi-logarithmic scale, which means some sort of saturation in the energy release rate. Fig. 7(d) shows the behavior of TJ displacement motion associated with the base extension and the peak height.

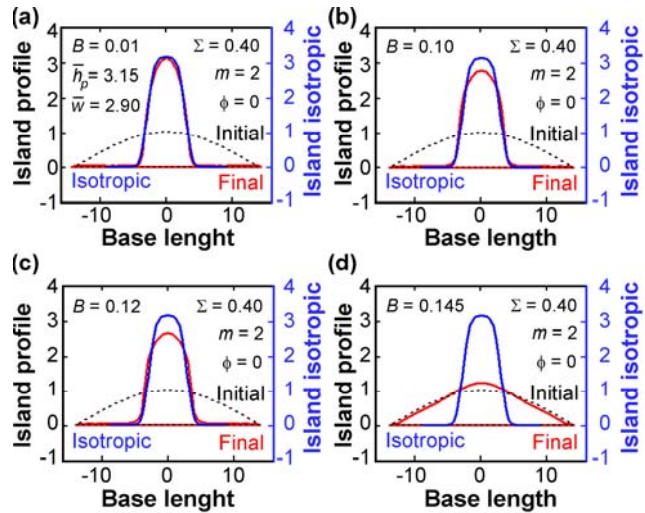

**FIG. 8. (Color online)** (a-c) Spontaneous formation of the SK singlets with an almost perfect flat wetting layer from a single crystal droplet having exposed to the anisotropic surface free energies having various intensities below the threshold level with the tilt angle of $\phi = 0^o \Rightarrow (1\bar{1}0)$. d) The formation of the *pyramidal shape* faceted islanding just above the onset of the anomalous instability regime without the wetting layer. Data: $\Sigma = 0.40$, $B = \{0.01 - 0.25\}$, $\bar{h}_p = 1$, $\beta = 28$, $\nu = 0.273$, $\bar{M}_{TJ} = 2$, $\lambda = 1$, $\bar{\delta} = 0.005$, $f_s = 1.2$ and $f_d = 1$.

In Fig. 8(a-d), we studied a series of fully developed SK singlets and the stabilization of the pyramidal-shape droplets below and above the threshold level of the anomalous anisotropy regime. Here, the SK singlets at the stationary states are supported by the thin wetting layer having a thickness of about $\Delta\bar{h} = 0.0314 \rightarrow 0.17\ nm$. In fact, we investigated a full range of anisotropy constants $B \subset [0.001 - 1]$, starting from very low values in the normal regime up to the high numbers in the anomalous region, where the surface stiffness takes negative values in certain well defined directions. In these simulation experiments, we utilized an ESED parameter of $\Sigma = 0.40$ with a tilt angle of $\phi = 0^o$. Fig. 8(c) corresponds to the case just below the threshold level, where one observes rather *dome-shape* island morphology having slight lower peak height, which is mostly takes places in isotropic systems and/or for the very low values of the surface free energy anisotropy constant as a typical example illustrated in Fig. 8(a,b).

In Fig. 9(a), we observed an extremely interesting morphological evaluation of a droplet, which is attached to the $(1\bar{1}0)$ top plane of a substrate for very high values of the surface free energy anisotropy constant in the anomalous regime such as $B \geq 1$. As one can see from Fig. 9(d), the overall system is in the transient nonequilibrium state.

Namely, the droplet shows apparently no shape changes other than the *monotonic decrease* in the peak height, which is in cooperated by the simultaneous base length extension that is a *linear function of time*. The temporal value of this extension amounts to 10%. On the other hand, the wetting angles at the TJ's show practically no departure from the initial value of about $\phi \cong 6.4^o$ other than some erratic fluctuations at the edges.

In Fig. 9(c), there is another very interesting case, which is the behavior of the capillary surface free energy change during the evolution process. The capillary surface free energy shows a linear increase with the normalized time, and has a same slope with the strain energy release. The strain energy release exhibits some erratic oscillations at $\bar{t} \leq 10^{-3}$, and otherwise it is about two orders of magnitude larger than the capillary free energy. In Fig. 9(d) one also observes steady decrease in the peak height in contrast to the base length, which shows linear time dependent increase, which is somewhat a slow evolution process. Nevertheless, that gives us a strong clue for the existence of the *Frank-van der Merwe* mode of thin film formation by the flatting or the base extension mechanism operating in the anomalous surface stiffness regime.

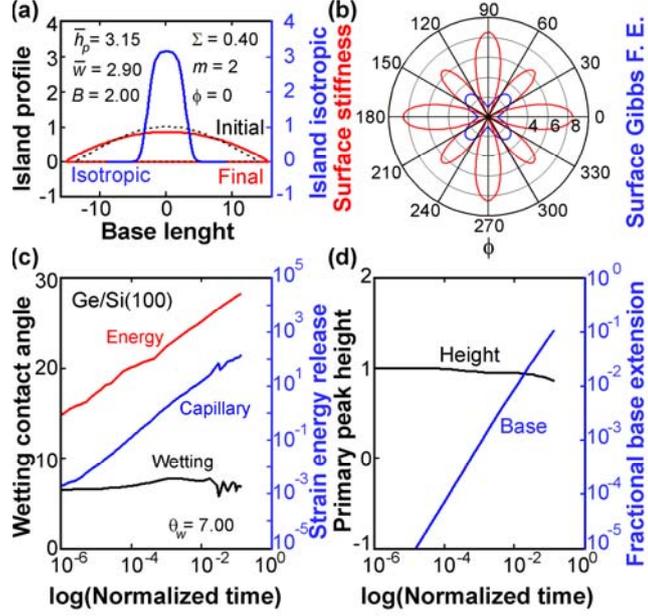

**FIG. 9. (Color online)** (a) The droplet at the zero tilt angle designated as $\phi = 0^o \Rightarrow <1\bar{1}0>$ shows no shape change even for the very large values of the anisotropy constant other than the slight spreading. (b) Surface free energy and stiffness are illustrated in polar plot, which shows $\langle 100 \rangle$ spikes having rather strong negative intensities that indicates the anomalous regime. (c) Evolution of the contact angle is shown on the left *y*-axis. On the right *y*-axis, the strain energy and surface free energy changes are given for Ge/Si($1\bar{1}0$) system. (d) Monotonic decrease in the peak height followed up by the substantial increase in the base length. Data: $\Sigma = 0.40$, $B = 2.0$, $\bar{h}_p = 1$, $\beta = 28$, $\nu = 0.273$, $\bar{M}_{TJ} = 2$, $\lambda = 1$, $\bar{\delta} = 0.005$, $f_s = 1.2$ and $f_d = 1$.

### c. Six-fold rotational symmetry in cubic structures: Zone axis $\langle 111 \rangle$

In this final section, we performed our computer simulations on a single crystal thin film droplet attached to one of those $\{1\bar{1}0\}$ *form of planes* as the top surface of a substrate described by the six-fold $[111]$ rotational symmetry axis in cubic crystal structures by using various tilt angles in the range of $\phi \supset \left(0^o - 60^o\right)$. Here zero tilt angle corresponds to any one of those six planes belong to the $\{1\bar{1}0\}$ form. From now on, only those experiments with an anisotropy constant of $B = 0.05 < 1/17$ will be reported. This anisotropy constant is specially chosen in order to amplify the effects since it is just below threshold level of the anomalous instability regime.

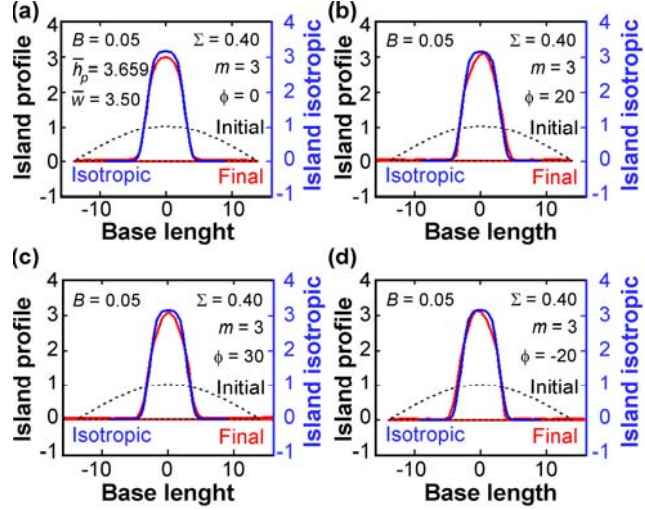

**FIG. 10. (Color online)** (a-d) Spontaneous formations of the SK singlets having different morphological appearances depending upon the tilt angles are presented for an anisotropy constant, $B = 0.05$, which is just below the anomalous threshold level of $B_{th} = 1/17$. Data: $\Sigma = 0.40$, $\bar{h}_p = 1$, $\beta = 28$, $\nu = 0.273$, $\bar{M}_{TJ} = 2$, $\lambda = 1$, $\bar{\delta} = 0.005$, $f_s = 1.2$ and $f_d = 1$.

Similarly, to illustrate the role of the surface free energy anisotropy constant alone on the morphological evolution of droplets, a series of simulations experiments are executed by using values $\{B \subset 0.001 - 1.0\}$ well below and above the anomalous threshold level of $B_{th} = 1/17$, for the special tilt angle of $\phi = 0^o$ that is the most interested substrate configuration in practice, namely $\{1\bar{1}0\}$. In Fig. 10, we illustrate the non-equilibrium stationary states of the single crystal droplets for a surface anisotropy constant of $B = 0.05$ after the morphological evolutions taking place at the four different tilt angles, which are specially selected configurations having six-fold rotational symmetries. Fig. 10(a-d) clearly shows that the tilt angle $\phi = 30^o$, which corresponds to the form of planes $\{11\bar{2}\}$ that belongs to $[111]$ zone axis, plays a special role, and acts as a quasi-reflection-symmetry axis for the final shapes of SK-islanding. That means the shape modifications and distortions on the SK-island singlets are symmetrically disposed in terms of their respective tilt angles relative to this orientation. The morphology of the SK-island obtained for the zero tilt angle as illustrated in Fig. 10(a) is very similar to the isotropic case, and it has a dome shape rounded top contour, which may be represented by a fourth degrees Gaussian type curve. On the other hand the SK-island presented in Fig. 10(c), which is obtained for the tilt angle $\phi = 30^o$ has a *sharply pointed top counter*, and looks like a cross section of a rounded hoot-shape islanding at 3D space as observed numerically by Golovin *et al.* **Error! Bookmark not defined.**[19] for [001] surface orientation in the literature.

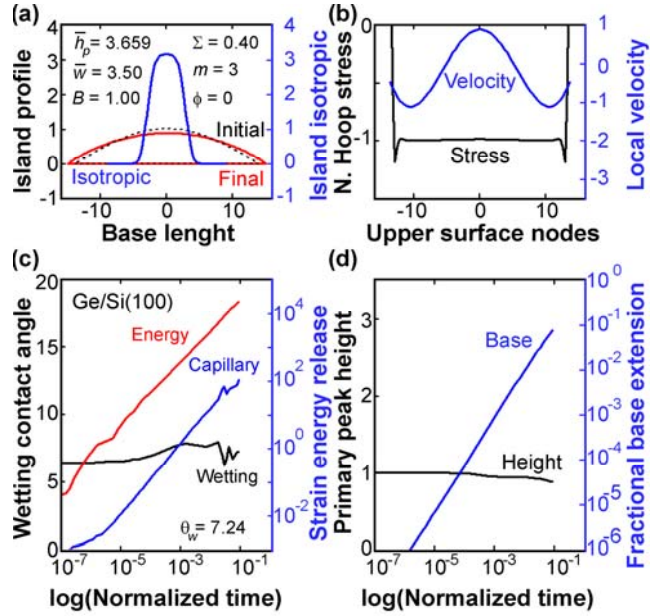

**FIG. 11. (Color online)** (a) The droplet at the zero tilt angle shows no shape change even for the very large values of the anisotropy constant in the anomalous regime. (b) Instantaneous velocity and the hoop stress distributions along the final droplet profile. (c) Evolution of the contact angle is shown on the left *y*-axis. On the right *y*-axis, the strain energy and surface free energy changes are given for the Ge/Si(100) system. (d) Time evolution of peak height and TJ displacement, which indicates the monotonic decrease in the peak height, and the steady increase in the base length, followed by some erratic variations in the wetting angles at the triple junctions.

In Fig. 11, we present the results of a simulation carried out with a very high value of the anisotropy constant, $B = 1$, and at the zero degree tilt angle. The applied anisotropy constant is well above of the lower limit or the threshold level of the anomalous instability regime designated by $B \geq 1/17$. Even with such a high value of *B*, the crystalline droplet shows no indication of the morphological changes in the original

Cosine-shape other than the adjustment of the triple junctions at the edges, and the substantial extension in the base length. This TJ adjustment as may be seen from Fig. 11(c) manifests itself by a major departure from the equilibrium configuration that is characterized by a wetting angle almost equal to zero degree. According to our observations, which are supported by our further simulation experiment that is not reported here, the base spreading increases with the anisotropy constant. Namely, the fractional changes in base length become: $\delta \ell / \ell = 0.01$ for $B = 0.01$ and $\delta \ell / \ell \geq 0.1$ for $B = 1.0$. That means no SK-islanding may be possible for the high anisotropy constants in the anomalous regime for the six-fold symmetries without having superimpose to the random undulations or white noise ripples over the droplet surfaces. In that case, one may talk about the *Frank-van der Merwe* mode of thin film formation by base-extension as a dominant morphological scenario instead of the *Volmer-Weber* islanding and/or the *Stranski-Krastanow* growth modes.

## IV. CONCLUSIONS

In summary, we studied the non-equilibrium stationary state morphologies of isolated thin solid droplets, below and above the anomalous surface stiffness threshold level, and obtained very interesting morphological scenarios. In this work, it is assumed that the evolution process is initiated by the nucleation route and a self-consistent 2D dynamical simulations having the free-moving boundary condition at the triple junction contour line were used. The anisotropic surface Helmholtz free energy, and the surface stiffness are all represented by the well accepted trigonometric functions.[41] While various tilt angles and anisotropy constants are considered during simulation experiments, the main emphasis were given on the two-fold, four-fold and six-fold rotational symmetries associated with the surface Helmholtz free energy topography in 2D space, to see their impacts on the final shapes of the SK-islanding. The following findings are observed as main features and/or characteristics of the morphological transitions taking place isothermally and spontaneously:

i. Simulations in the two-fold and four-fold rotational symmetries revealed that for a given tilt angle, there are two well-defined domains in the surface free energy anisotropy constant. The first domain is below the threshold level of anomalous regime and characterized by the morphological transition of a droplet into the SK island formation embedded in a wetting layer platform. This transition may occur as both singlets and/or doublets depending upon the tilt angle. In the second domain (*i.e.*, the anomalous regime), a partial stabilization of the initial shape of the droplet occurs if the anisotropy constant is just above the threshold level. Otherwise the *Frank-van der Merwe* mode of thin film formation by base-extension takes place by following a linear kinetic law. Conversely, in the case of six-fold symmetry, the SK singlet islanding takes place regardless of the tilt angle as long as one stays in the normal stability regime.

ii. Six-fold rotational symmetry associated with the $\{1\bar{1}0\}$ form of planes as the top substrate planes with zero tilt angle in the anomalous regime has a unique property of showing absolute nonequilibrium stability of the original droplet even for very high values of the anisotropy constant, with the major base extension by almost keeping the original wetting angles at TJ-edges. This behavior implies the formation of *Frank-van der Merwe* mode of thin film as dominant scenario during the strain-heteroepitaxial growth if one have high anisotropies in the surface free energies.

iii. In general for a given anisotropy constant, the tilt angle may have profound effects on the morphology of the SK islanding. Arbitrarily selected tilt angle may cause distortion on the peaking shape rather than producing sharp faceting as illustrated in this paper. Similarly, there are certain orientations exist, which may result doublet formation rather than the singlet islanding that is especially the typical case for the zero tilt angle.

iv. A careful examination of the simulation results also demonstrated that SK-singlet always prefers to those vicinal planes (*i.e.*, cusps), which have the lowest surface free energy in a given set of planes associated with the rotational symmetry axis designated as the zone axis. On the other hand, the SK-doublets prefer to base on those substrate planes, which have higher surface free energies. This could be easily understood by looking at the surface stiffness map, which shows that the cusp corresponds to maxima in the stiffness plot, and the maxima in the surface free energy between cusps produces minima in the surface stiffness mapping. Since normalized curvature in the governing equation is augmented by the stiffness the capillary potential is totally controlled by the stiffness along the SK-profile but not by the surface free energy itself.

**ACKNOWLEDGMENTS**

The authors extend their thanks to Oncu Akyildiz of METU for his valuable comments on the paper. This work was partially supported by the Turkish Scientific and Technological Research Council, TUBITAK through a research Grant No.107M011.

**REFERENCES**

[1] C. W. Snyder, B. G. Orr, D. Kessler and L. M. Sander, Phys. Rev. Lett. **66** (23), 3032 (1991).

[2] D. J. Eaglesham and M. Cerullo, Phys. Rev. Lett. **64** (16), 1943 (1990).

[3] D. Leonard, M. Krishnamurthy, C. M. Reaves, S. P. Denbaars and P. M. Petroff, Appl. Phys. Lett. **63** (23), 3203 (1993).

[4] G. Wang, S. Fafard, D. Leonard, J. E. Bowers., J. L. Merz and P. M. Petroff, Appl. Phys. Lett. **64** (21), 2815 (1994).

[5] J. Berrehar, C. Caroli, C. Lapersonne-Meyer and D. C. Schott, Phys. Rev. B **46** (20), 13487 (1992).

[6] R. M. Tromp, F. M. Ross, and M. C. Reuter, Phys. Rev. Lett. **84**, 4641 (2000).

[7] T. O. Ogurtani, J. Apply. Phys. **106** (5), 01-Sep-09 scheduled (2009).

[8] B. J. Spencer, P. W. Voorhees and S. H. Davis, Phys. Rev. Lett. **67** (26), 3696 (1991).

[9] B. J. Spencer, P. W. Voorhees and S. H. Davis, Phys. Rev. B **47** (15), 9760 (1993).

[10] B. J. Spencer, Phys. Rev. B **59** (3), 2011 (1999).

[11] C. H. Chiu and H. Gao, in Thin Films: Stresses and Mechanical Properties V, edited by S.P. Baker *et al.*, MRS Symposia Proceedings No. 356 (Materials Research Society, Pittsburgh), p. 493 (1995).

[12] L. B. Freund and F. Jonsdottir, J. Mech. Phys. Solids **41** (7), 1245 (1993).

[13] W. T. Tekalign and B. J. Spencer, J. Appl. Phys. **96** (10), 5505 (2004).

[14] W. T. Tekalign and B. J. Spencer, J. Appl. Phys. **102**, 073503 (2007).

[15] Y. W. Zhang and A. L. Bower, J. Mech. Phys. Solids **47**, 2273 (1999).

[16] Y. W. Zhang, Appl. Phys. Lett. 75 (2), 205 (1999).

[17] R. Krishnamurthy and D. J. Srolovitz, J. Appl. Phys. **99**, 043504 (2006).

[18] N. V. Medhekar and V. B. Shenoy, J. Appl. Phys. **103**, 063523 (2008).

[19] M. S. Levine, A. A. Golovin, S. H. Davis and P. W. Voorhees, Phys. Rev. B **75**, 205312 (2007).

[20] Y. W. Zhang and A. L. Bower, Appl. Phys. Lett. **78** (18), 2706 (2001).

[21] P. Liu, Y. W. Zhang, and C. Lu, Phys. Rev. B **68**, 035402 (2003).

[22] A. A. Golovin, M. S. Levine, T. V. Savina and S. H. Davis, Phys. Rev. B **70**, 235342 (2004).

[23] H. R. Eisenberg and D. Kandel, Phys. Rev. B **71,** 115423 (2005).

[24] T. O. Ogurtani, J. Chem. Phys. **124** (13), 144706 (2006).

[25] T. O. Ogurtani and E. E. Oren, Int. J. Solids Struct. **42**, 3918 (2005).

[26] T. O. Ogurtani, O. Akyildiz and E. E. Oren, J. Appl. Phys. **104**, 013518 (2008).

[27] W. Wang and Z. Suo, J. Mech. Phys. Solids, 45, 709 (1997).

[28] L. E. Murr, Interfacial Phenomena in Metals and Alloys (Addison-Wesley, London,1975) p.186.

[29] Y. Pang and R. Huang,, Phys. Rev. B **74,** 075413 (2006).

[30] G. Beer and J. O. Watson, *Introduction to finite and boundary element methods for engineers* (Wiley, New York, 1992) p.151.

[31] T. O. Ogurtani and O. Akyildiz, J. Appl. Phys. 104, 023521; 023522 (2008).

[32] T. O. Ogurtani and O. Akyildiz, Int. J. Solids Struct. **45**, 921 (2008).

[33] C. W. Gear, *Numerical Initial Value Problems in Ordinary Differential Equations* (Prentice Hall, New Jersey, 1971), p.151.

[34] J. Pan and A. C. F. Cocks, Acta Metall. Mater. **43** (4), 1395 (1995).

[35] R. V. Kukta and R B. Freund, J. Mech. Phys. Solids **45**, 1835 (1997).

[36] J. R. Rice and D. C. Drucker, Int. J. Fract. Mech. **3**, 19 (1967).

[37] H. Gao, Int. J. Solids Struct. **28** (6), 703 (1991).

[38] J. D. Eshelby, In: M. F. Kanninen, W. F. Adler, A. S. R. Rosenfield, and R. Jaffe, (Ed.), Inelastic Behavior of Solids (McGraw-Hill, New York, 1970) p. 90.

[39] H. Gao, J. Mech. Phys. Solids **42**, 741 (1994).

[40] M. Krishnamurthy, J. S. Drucker, and V. Venables, J. Appl. Phys. 69 (9), 6461 (1991).

[41] T. O. Ogurtani. Phys, Rev. B 74, 155422 (2006).